\documentclass[%
 amsmath,amssymb,
 aps,
 pra,
twocolumn
]{revtex4-2}

\usepackage{graphicx}%
\usepackage{multirow}%
\usepackage{amsmath,amssymb,amsfonts}%
\usepackage{amsthm}%
\usepackage{mathrsfs}%
\usepackage[title]{appendix}%
\usepackage{xcolor}%
\usepackage{textcomp}%
\usepackage{booktabs}%
\usepackage{algorithm}%
\usepackage{algorithmicx}%
\usepackage{algpseudocode}%
\usepackage{listings}%
\usepackage{verbatim}   
\usepackage{physics}
\usepackage{soul}

\usepackage{dcolumn}
\usepackage{bm}

\definecolor{darkgreen}{HTML}{006400}
\usepackage[normalem]{ulem}

\newcommand{\affilATI}{Vienna Center for Quantum Science and Technology (VCQ), Atominstitut, TU Wien, Vienna, Austria}

\begin{document}

\title{Universal non-thermal fixed point for quasi-1D Bose gases}

\author{Qi Liang}
\affiliation{\affilATI}

\author{RuGway Wu}
\email{rugway.wu@tuwien.ac.at}
\affiliation{\affilATI}

\author{Pradyumna Paranjape}
\affiliation{\affilATI}

\author{Ben Schittenkopf}
\affiliation{\affilATI}

\author{Chen Li}
\affiliation{\affilATI}

\author{J\"org Schmiedmayer}
\affiliation{\affilATI}

\author{Sebastian Erne}
\email{sebastian.erne@tuwien.ac.at}
\affiliation{\affilATI}

\begin{abstract}
Spatio-temporal scaling dynamics connected to non-thermal fixed points has been suggested as a universal framework to describe the relaxation of isolated far-from-equilibrium systems. Experimental studies in weakly-interacting cold atom systems have found scaling dynamics connected to specific attractors. In our experiments, we study a quantum gas of strongly interacting $^6$Li$_2$ Feshbach molecules, brought far out of equilibrium by imprinting a white-noise phase profile. The observed relaxation follows the same universal dynamics as for the previously observed formation of the order parameter in a shock-cooled gas of weakly interacting $^{87}$Rb atoms. Our results point to a single universal fixed point with a large basin of attraction governing the relaxation of quasi-1D bosonic systems, independent of their specific initial conditions and microscopic details.
\end{abstract}

\maketitle

Understanding non-equilibrium dynamics of quantum many-body systems remains a formidable challenge \cite{nonequilibriumRev_2011}. Often separate descriptions are developed for each system, and a general understanding and classification, especially for far-from-equilibrium systems, is still elusive. For systems close to thermal equilibrium, such a systematic classification into well-defined universality classes \cite{RevModPhys_1977_criticalPhenomena} enabled the identification and deep understanding of phenomena based on their critical exponents.
In recent years, growing evidence of spatio-temporal scaling dynamics emerged even for far-from-equilibrium systems, both from theoretical considerations and experimental observation (see e.g.~\cite{2019_nonthermal_fixed_point_review}). 

On the theoretical side it was suggested that far-from-equilibrium evolution of isolated quantum systems follows attractor solutions governed by so-called Non-thermal fixed points (NTFPs) \cite{Berges_2008_nonthermal_fixed_point}. In its vicinity, the dynamics can be simplified and characterized by a set of universal exponents, leading to the rapid loss of information on the details of the initial state. There is growing theoretical and numerical evidence that a large class of microscopically different systems follow such universal attractor solutions, which can encompass both relativistic and non-relativistic models \cite{Pineiro_Orioli_2015_self_similar_dynamics}, ranging from reheating in the early universe \cite{Gasenzer2012inflationNTFP,Chatrchyan2021reheatingNTFP} to QCD \cite{Berges_2019_prescaling}, from heavy ion collisions \cite{Berges_2015_Universality} to ultra-cold quantum gases \cite{Walz_2018_largeN_kineticTheory,Heller2024PrescalingRelax}. This suggests that, in analogy to second-order phase transitions and critical phenomena, these can be classified according to their exponents and scaling functions.

First direct experimental observations were made in quasi-1D spinor Bose-Einstein condensates (BEC) \cite{Pruefer_2018_scaling_spinor_BEC} and quench-cooled Bose gases \cite{Erne_2018_universal_scaling}, followed by additional studies within various cold-atom systems \cite{Glidden_2021_bidirectional_scaling,gazo2023universal,martirosyan_2023_driven_boxTrapped_bose_gas,huh_2024_quenched_ferromagnetic,martirosyan_2024_universalspeedlimitspreading}. Finding and characterizing such non-equilibrium universality classes, each connected to a specific NTFP \cite{huh_2024_quenched_ferromagnetic,lannig2023TwoNTFP}, would be the first step in the direction of a universal description of non-equilibrium phenomena. However, to date, the hallmark of universality, that is, a direct experimental observation of microscopically different systems approaching the same NTFP, is still absent. 

\begin{figure}[!b]
\center
\includegraphics[width=\columnwidth]{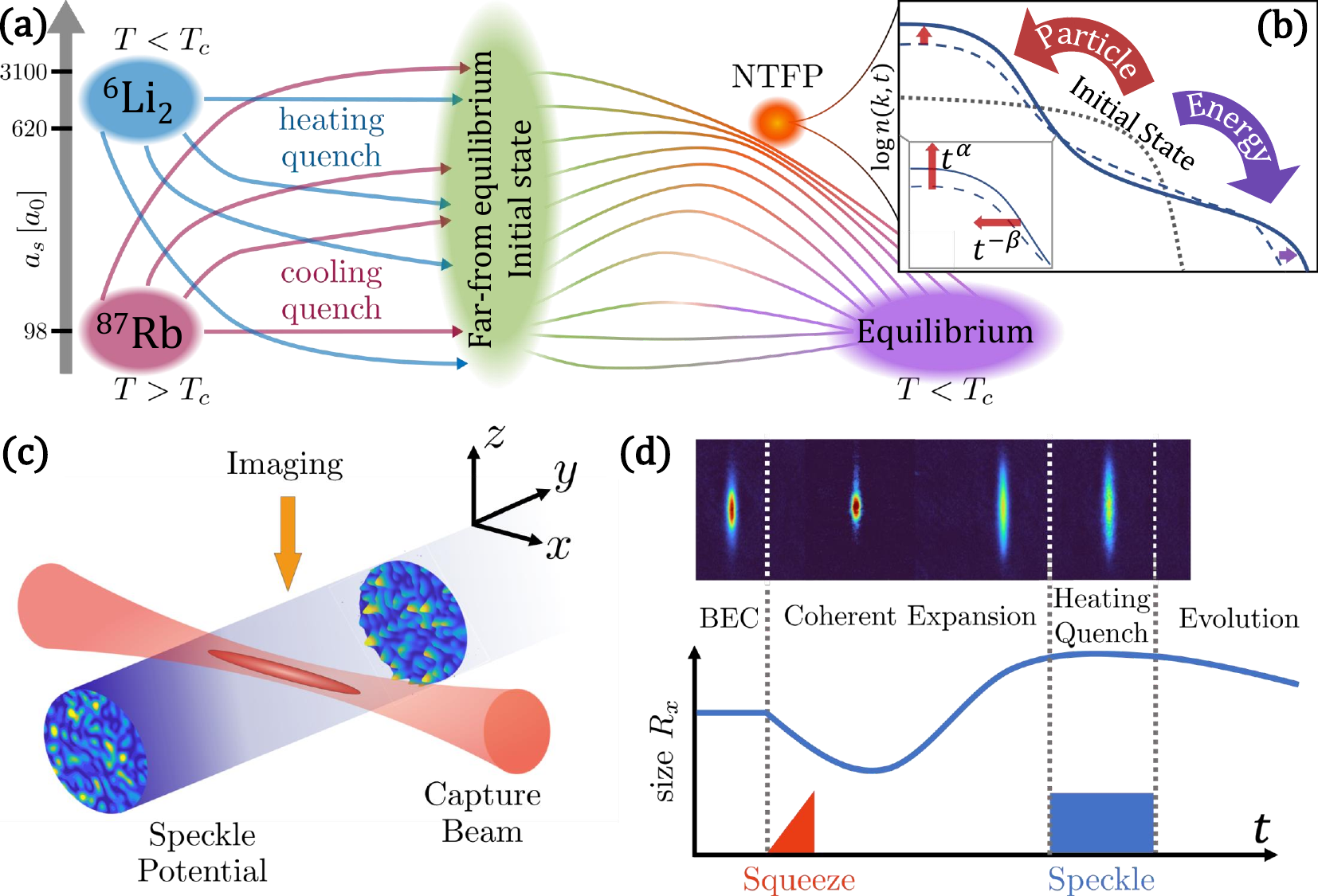}
\caption{\textbf{Universal scaling and experimental approach:}
(\textbf{a}) Isolated quantum systems quenched to far-from-equilibrium state can approach attractor solutions called non-thermal fixed points (NTFPs) in the course of relaxation, leading to (\textbf{b}) universal self-similar scaling dynamics during the two-way transport process. A few particles will be transported to high-momentum modes with excessive energy, while a fraction of particles will be transported to low-momentum modes. (\textbf{c}) Qualitative sketch of the experimental arrangement. Imaging axis $z$ is the gravitational direction. (\textbf{d}) The schematic of the quench sequence. An additional attractive beam is implemented along the $y$ axis to squeeze and release the cloud before pulsing the speckle potential at its maximum size, such that the breathing motion could be suppressed.}
\label{fig:setup_and_profiles}
\end{figure}

We experimentally study the relaxation dynamics of a strongly perturbed elongated gas of bosons consisting of Feshbach molecules of $^6$Li \cite{jochim2003BEC}, varying the inter-molecular scattering length $a_{dd}$ over a wide range. We find two distinct time periods within which the system exhibits self-similar scaling dynamics. Furthermore, the universal exponents and function found for the first period coincide with the results \cite{Erne_2018_universal_scaling} for weakly interacting shock-cooled quasi-one-dimensional $^{87}$Rb. Using the random defect model \cite{Schmidt_2012_RDM}, we establish a quantitative connection between universal evolution rates and soliton-like excitations, while also revealing the role of dimensionality. The combined results indicate a single universal NTFP that governs the far-from-equilibrium condensation dynamics of quasi-1D bosonic systems, regardless of interaction strength, as depicted in Fig.~\ref{fig:setup_and_profiles}(a),(b). Our results open new avenues for investigation: from analogue simulations of strongly interacting systems far-from-equilibrium to connections with, e.g., Generalized Hydrodynamics (GHD) \cite{Fujii2024GHDAttractor} and potential classifications of far-from-equilibrium universality classes based on non-integrable corrections \cite{durnin2021brokenIntegrability}.

Our experiment starts with a cigar-shaped, nearly pure molecular Bose-Einstein condensate (mBEC) of $^6$Li$_2$ Feshbach molecules in an elongated potential with trapping frequencies $(f_x,f_y,f_z) = {(16,100,100)}\,\mathrm{Hz}$. A schematic of the experiment is shown in Fig.~\ref{fig:setup_and_profiles}(c). The system is brought to a far—from—equilibrium state through a combination of a coherent expansion and a pulsed optical speckle potential \cite{Lye_2005_BEC_randomPotential} imprinting a random phase pattern on the order parameter (see Supplementary for details on the preparation and quench protocol).

The two-step quench sequence, sketched in Fig.~\ref{fig:setup_and_profiles}(d), enables us to minimize unwanted mean-field dynamics of the bulk density (breathing excitation) caused by the longitudinal harmonic confinement. We apply a speckle pulse of duration $1.5 \, \mathrm{ms}$, which is strong enough to completely scramble the phase of the order parameter and follow its reformation.

This \emph{quench} induces a highly occupied white spectrum of excitations \cite{Nagler_2022_specklePulse}, leading to a strongly broadened momentum distribution along the longitudinal $x$ direction. The relaxation of this far-from-equilibrium state to a stationary state is then probed by holding it for a variable evolution time $t>0$ in the trap. The momentum distribution of the molecules is measured by switching off the trapping potential and employing matter wave focusing (see \cite{Murthy_2014_focusing} and Supplementary).

\begin{figure}[!t]
\center
\includegraphics[width=0.9\columnwidth]{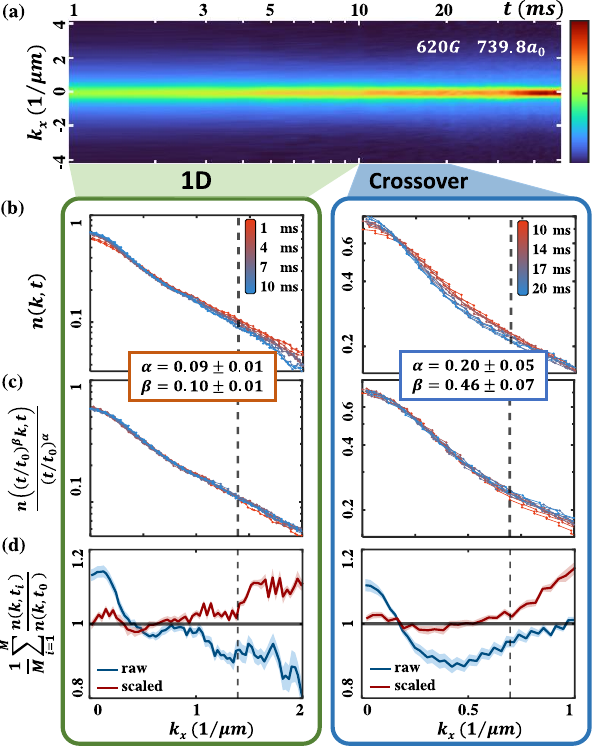}
\caption{\textbf{Characteristic time evolution:} (\textbf{a}) Time carpet for the longitudinal momentum distribution $k_x$, measured with $B=620\,\mathrm{G}, a_{dd}=740a_{0}$. Colorbar represents the normalized density. Each distribution is centered by fitting a concentric double Gaussian profile. (\textbf{b}) The raw and (\textbf{c}) scaled momentum profiles within the ``1D''(left) and the ``Crossover''(right) scaling windows. (\textbf{d}) The mean ratio of raw (blue) and scaled (red) profiles divided by the reference profile $n(k,t_0)$. Scaling cutoff is marked by vertical dashed lines.
}
\label{fig:Evolution}
\end{figure}

A typical time evolution of the momentum distribution is shown in Fig.~\ref{fig:Evolution}(a) for the case of $N \simeq 7500$ molecules and an inter-molecule $s$-wave scattering length $a_{dd} =$ 740$a_0$, with the Bohr radius $a_0$. The isolated system relaxes to a stationary state within the first $\approx 40 \, \mathrm{ms}$, during which we find transport of particles towards the infrared (smaller momenta) leading to the re-emergence of a uniform condensate.

In the vicinity of NTFP, the system is expected to exhibit universal self-similar scaling dynamics of correlations, such as, e.g., the momentum distribution
\begin{equation}
    n(k,t) =  \left(t/t_0\right)^{\alpha}\,f_\mathrm{S}\left[ \left( t/t_0 \right)^{\beta}\,k \right] ~.\label{equation:scaling_evolution}
\end{equation} 
Here $t_0$ is an arbitrary reference time within the period of self-similar scaling. Therein, the evolution of the system is characterized only by a set of time-independent scaling exponents $\alpha,\beta$ and functions $f_S$.

We demonstrate in Fig.~\ref{fig:Evolution}(b)-(d) that we can indeed identify two time periods during which the system exhibits self-similar scaling dynamics according to Eq.~\ref{equation:scaling_evolution}. Within each period, we find a distinct pair of exponents $\alpha,\beta>0$ for which the evolving longitudinal momentum distribution $n(k_x,t)$ can efficiently be scaled to collapse to a single time-independent function $f_\mathrm{S}(k)=n(k,t_0)$. 

The scaling exponents are determined via a combined maximum likelihood fit of Eq.~\ref{equation:scaling_evolution} for all reference times $t_0$ within the scaling period (see Supplementary for details on the error estimation). We find $\alpha=0.09 \pm 0.01$ and $\beta=0.10 \pm 0.01$ for the first (``1D'') and $\alpha=0.20 \pm 0.05$ and $\beta=0.46 \pm 0.07$ for the second (``crossover'') scaling window. Note that in general we find good agreement with the predicted scaling behavior of spatially averaged observables, i.e.~moments $\bar{M}_n(t) \sim \int \mathrm{d}k \, |k|^n n(k,t)$ of the distribution \cite{Erne_2018_universal_scaling,Berges_2019_prescaling} (see Figs.~\ref{fig:scalingWindow},\ref{fig:globalvariable} and Supplementary for details).

The observed self-similarity according to Eq.~\ref{equation:scaling_evolution} constitutes an enormous reduction in complexity for the time-evolution of $n(k,t)$, as it does not depend separately on $k$ and $t$ but only on the appropriate product determined by the scaling exponents $\alpha$ and $\beta$. Moreover, any additional relevant parameter to the dynamical evolution would result in a further dependence of Eq.~\ref{equation:scaling_evolution} on this quantity, potentially via a new scaling exponent.

\begin{figure}[t]
\center
\includegraphics[width=0.9\columnwidth]{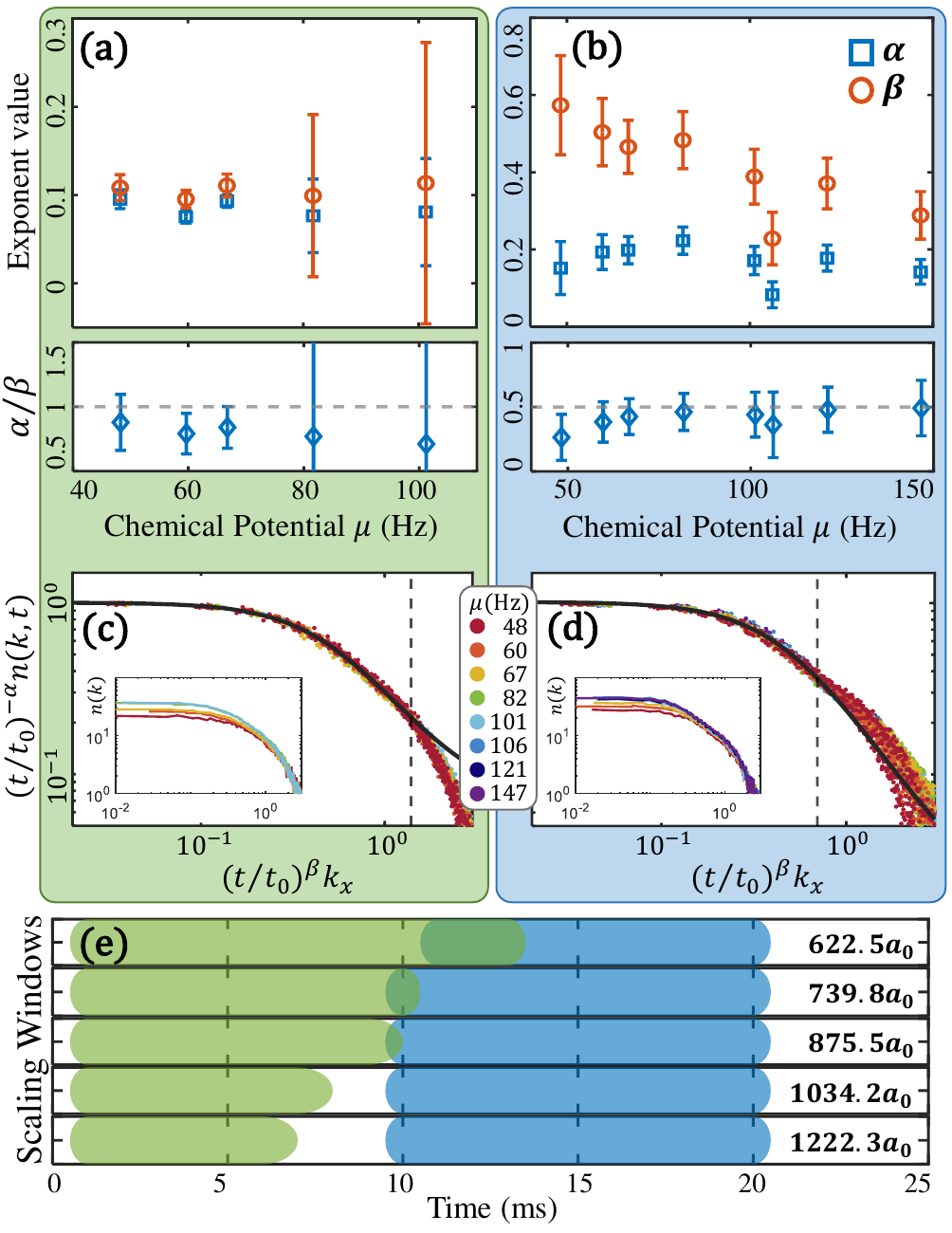}
\caption{\textbf{Scaling properties and time windows:} Scaling exponents and the ratio of (\textbf{a})1D and (\textbf{b})crossover windows versus chemical potential ($\mu$ calculated from initial density). As the 1D window duration decreases with increasing interaction, scaling exponents can no longer be determined for $\mu>100\,\mathrm{Hz}$ due to limited statistics. Universal scaling function for (\textbf{c})1D and (\textbf{d})crossover window. All momentum distributions (dots) within the scaling window collapse to the universal function $f_s=(1+k^{\zeta})^{-1}$ (solid line) with (\textbf{c})$\zeta=1.604\pm0.011$ and (\textbf{d})$\zeta=1.766\pm0.014$. The vertical dash lines mark the scaling cutoff of each window. The insets plot the initial momentum distributions for datasets of different interaction strengths. (\textbf{e}) The duration of scaling windows at different interaction. For weaker interaction, the two scaling windows have an overlap, during which both set of scaling exponents are valid within their momentum cutoff. Towards stronger interactions a time gap widens between them, confirming mutual independence of the duration of the two scaling windows.}
\label{fig:Exponents}
\end{figure}

In order to test such an insensitivity of the fixed point to microscopic details, we probe the self-similar scaling evolution for mBECs at different interaction strengths, by varying the molecule-molecule $s$-wave scattering length $a_\mathrm{dd}$ via the Feshbach resonance. Fig.~\ref{fig:Exponents} depicts a summary of the scaling analysis for varying interactions, from $620 a_0$ to $2000 a_0$. 
For later convenience, we characterize the tunable interaction via the local chemical potential $\mu = n_0 g_\mathrm{1D}$ directly following the quench. Here, $n_0$ is the 1D peak density of the initial state, and $g_\mathrm{1D}=16 \hbar^2a_{dd}/(3mR_{\perp}^2)$ is the effective 1D interaction constant, with the radial size of the cloud $R_{\perp}$ (fitting a Thomas-Fermi profile).

The exponents $\alpha \! \approx \! \beta \! \approx \! 0.1$ for the earlier scaling window (Fig.~\ref{fig:Exponents}(a)) are found to be robust over a wide range of interaction strengths with a well-preserved ratio of $\alpha/\beta \approx 1$. This is in accordance with the theoretical prediction $\alpha = d\beta$ for particle-conserving transport within the scaling region, where $d=1$ is the effective dimensionality of the system. We therefore refer to it as the ``1D-scaling'' window.

For the second scaling window (Fig.~\ref{fig:Exponents}(b)) the exponents found decrease for stronger interactions, while quickly approaching a constant ratio of $\alpha/\beta \approx 0.5$. Interestingly, this predicts the emergence of a conserved quantity $\sim 1/\sqrt{k}$ within the scaling region, being transported toward lower momenta. Consequently, particle transport towards the infrared is reduced, with the particle number within the scaling region decreasing $\sim t^{\alpha-\beta} \approx t^{-\alpha/2}$. A possible explanation for this could be coupling from the radial directions. We hence refer to the second time period as the ``crossover-scaling'' window.

We further demonstrate for both scaling periods that the shape of the scaling function $f_\mathrm{S}$ is universal and independent of the scattering length $a_{dd}$. This enables us to collapse all measurements that lie within the 1D-scaling or crossover-scaling window to a single universal function, respectively (see Fig.~\ref{fig:Exponents}(c),(d)). In other words, we found scaling parameters $\alpha(\mu), \beta(\mu)$ and functions $f_S$ such that, for any interaction strength, the non-equilibrium evolution of the system in the vicinity of the fixed point is fully described by Eq.~\ref{equation:scaling_evolution}.

\begin{figure*}[!t]
\center
\includegraphics[width=0.9\textwidth]{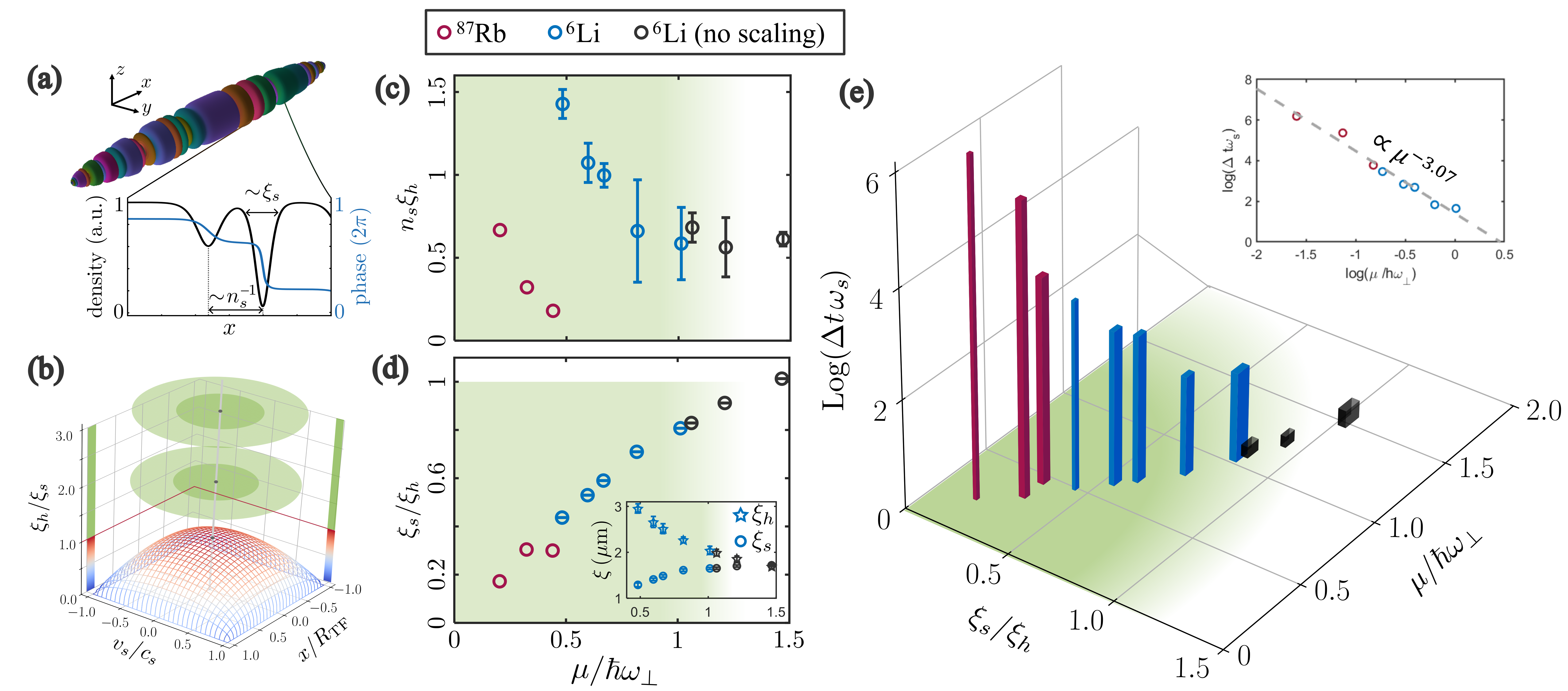}
\caption{\textbf{Generalized solitonic defect states and basin of attraction:} 
(\textbf{a}) Generalized solitonic defect state. The state is a product state of generalized solitonic defects (allowing for an arbitrary width $\xi_s$) with density $n_s$. Associated with the solitonic defect is a localized density suppression (radial width in the colorplot) and a phase slip $|\Delta \theta| \leq \pi$ (encoded in color). (\textbf{b}) The defect width of the GSD state in the solitonic phase space. Stable solitonic solutions are located at $\xi_s \geq \xi_h$ (curve surface below the red line) whereas GSD states have an average width $\xi_s < \xi_h$. The two green surfaces indicate the location of the $^{87}$Rb and $^6$Li$_2$ scaling data. Fit results of initial defect (\textbf{c}) density and (\textbf{d}) size for the GSD state for $^6$Li$_2$ (blue) and $^{87}$Rb \cite{Erne_2018_universal_scaling} (red) as a function of $\mu/\hbar \omega_\perp$, characterizing the degree of integrability breaking for different interactions. Note the dimensionless units in terms of the healing length $\xi_h$. The green shaded area indicates the expected region of self-similar scaling (\textbf{e}) NTFP basin of attraction. Dimensionless duration of the 1D-scaling period in the reduced parameter space for $^{87}$Rb (red) and $^6$Li$_2$ showing a scaling window (blue) and no scaling (black) within the experimental precision. The data indicates the fixed point near $(0,0)$. The green area indicates again the expected region of scaling. The inset plot the $\mu/\hbar\omega_{\perp}$ versus dimensionless duration. The dashed gray line shows the power law fit for both $^{87}$Rb and $^6$Li$_2$ data combined. }
\label{fig:Rb_comparison}
\end{figure*}

Moreover, in Fig.~\ref{fig:Exponents}(e) we depict the duration $\Delta t$ of the two scaling windows, estimated via the maximum likelihood fit. The measured extent of the crossover-scaling window is rather independent of $a_{dd}$ and its beginning is here determined by the transverse confinement (see Fig.~\ref{fig:scalingCossoverWindow} and Supplementary). For a different quench sequence that replaces the coherent expansion via squeezing with an incoherent expansion using another speckle pulse, its beginning is shifted (see Fig.~\ref{fig:scaling_vertical} and Supplementary). This further corroborates that the second scaling window is affected by the radial directions. Note, however, that we cannot reliably measure the in-trap evolution of the radial momentum distribution, due to the rapid interaction-dominated expansion during matter-wave focusing (see Fig.~\ref{fig:transversMomentum}). We therefore concentrate in the following on the ``1D'' scaling window, which shortens with increasing interaction strength and is absent for $\mu / h \gtrsim 100 \, \mathrm{Hz}$ or the incoherent double speckle sequence.

Finally, we compare our results to Ref.~\cite{Erne_2018_universal_scaling}, where equivalent scaling exponents $\alpha \approx \beta \approx 0.1$ were found during the relaxation following a cooling quench of $^{87}$Rb ($a_s \simeq 98 a_0$, $\omega_\perp=3.3\,\mathrm{kHz}$). The momentum distribution directly following the quench is for both experiments in good agreement with a model of randomly distributed solitonic defects \cite{Schmidt_2012_RDM} (see Fig.~\ref{fig:solitonDensityFit} and Supplementary). A schematic of the order-parameter for such a generalized solitonic defect (GSD) state is depicted in Fig.~\ref{fig:Rb_comparison}(a). The ensemble is conveniently described via a probability distribution $P(x,v_s,\xi_s)$ in the (extended) soliton phase-space Fig.~\ref{fig:Rb_comparison}(b). For a uniform ensemble with defect size $\xi_s$, the momentum distribution has a Lorentzian shape $n(k) \sim [1 + k n_s]^{-1}$ at small momenta, determined by the defect density $n_s$, and exhibits a distinct non-thermal decay $n(k) \sim \operatorname{exp}(- \xi_s k)$ at high momenta, connected to the localized density suppression of a solitonic defect. Note that for stable soliton solutions of the non-linear Schr{\"o}dinger equation, we have $\xi_s \geq \xi_h$, where $\xi_h = \hbar / \sqrt{2 m \mu}$ is the so-called healing length of the system.

The fit results for both experiments are presented in Figs.~\ref{fig:Rb_comparison}(c),(d) (see also table \ref{Table:Dataset_overview}). In order to fully compare the new results with the previous $^{87}$Rb experiment we consider dimensionless parameters $\xi_s / \xi_h$ and $n_s \xi_h$, which eliminates their trivial dependence for varying interactions $\mu$. Furthermore, we consider the ratio $\mu/\hbar \omega_\perp$ to parametrize the different interactions and trap geometries. This characterizes the degree of integrability breaking, i.e.~the 1D-ness of the gas, and allows us to directly compare the quasi-1D regime of \cite{Erne_2018_universal_scaling} with the elongated 3D regime with tunable inter-molecule interactions $a_{dd}$ considered here. Note, however, that while $\mu/\hbar \omega_\perp$ is commonly associated with the dynamical stability of dark solitons \cite{Muryshev_1999_stability_standing_matter_wave,Frantzeskakis_2010_dark_soliton}, a direct comparison to the theoretical and numerical critical values for soliton stability is not applicable for our far-from-equilibrium GSD states.

We observe an increasing (decreasing) defect width (density) for increasing interactions. For our quench protocol, their absolute values are dominated by the details of the optical speckle potential, in particular the fixed speckle-size (see Supplementary). In contrast, for the cooling quenches in \cite{Erne_2018_universal_scaling} no such external length scale is imposed by the quench.
We find an effective defect width $\xi_s / \xi_h \lesssim 1$, to be a necessary signature for the system to undergo universal self-similar coarsening dynamics with scaling exponents $\alpha=\beta\approx0.1$ (indicated by the green shaded area in Figs.~\ref{fig:Rb_comparison}(c),(d)). Interestingly, for these states the averaged defect width lies below the minimal value for stable dark solitonic solutions (see Fig.\ref{fig:Rb_comparison}(b)). The observed self-similar coarsening dynamics can equivalently be phrased as a dilution of the GSD state, i.e.~a decreasing defect density $n_s \sim t^{-\beta}$. Hence, as we are interested in this infrared scaling dynamics, we define $\omega_\mathrm{s} = 2\hbar n_\mathrm{s}^2 / m$, for the initial state defect density which sets the typical timescale of the evolution.

We present in Fig.~\ref{fig:Rb_comparison}(e) a summary of the dimensionless scaling time $\Delta t \omega_\mathrm{s}$ for all measurements in the $(\xi_s / \xi_h \, , \, \mu/\hbar \omega_\perp)$ plane, indicating the basin of attraction of the fixed point, located near $(0,0)$. Since it is expected, due to kinematic restrictions, that the scaling exponents vanish for a strictly 1D system \cite{gresista2022dimensional}, the limit $\mu/\hbar \omega_\perp \to 0$ has to be understood in such a way that the system remains in the quasi-1D regime. The consistency between the two experiments, i.e.~common scaling exponents $\alpha,\beta$ and universal functions $f_S$ (see Supplementary for details), strongly indicates that the exact microscopic details of the non-integrable contributions are irrelevant for the universal properties of the fixed point.


Our work experimentally demonstrates self-similar scaling evolution in momentum for the first time in a far-from-equilibrium composite Bose gas ($^6$Li$_2$ Feshbach molecules) with \textit{strong} inter-particle interactions.
Furthermore, we were able to relate our findings to an independent experiments with \textit{weakly} interacting bosonic $^{87}$Rb atoms.  Together, the two experiments span over more than an order in magnitude in interaction strength ($a_s = 95 - 2000 a_0$), and different quench protocols (shock cooling a thermal gas vs scrambling the order parameter of quantum gas) through concepts of universality far-from-equilibrium. 
Together, the consistent observation of self-similar scaling (i.e.~$\alpha, \beta, f_S$) signals the existence of a universal NTFP, which so far has neither been predicted nor fully understood theoretically.

The observation that the strength of the interaction (here $a_{dd}$) does not influence the universal properties of the quasi-1D fixed point opens the way towards simulating far-from-equilibrium dynamics/properties of strongly interacting quantum many-body systems through, e.g., weakly-interacting quantum analogues operating on experimentally more favorable regimes. Further, the relevance and observed insensitivity to the details of integrability breaking corrections opens fascinating avenues for investigation; from connections to generalized hydrodynamics \cite{Panfil_2023_GHD,GHD_PRX} to potential distinctions/definition of far-from-equilibrium universality classes based on non-integrable corrections.

\vspace{10pt}
We thank Zoran Hadzibabic, Thomas Gasenzer, Maximilian Pr\"{u}fer, and Frederik M{\o}ller for helpful discussions. This work was supported by the Austrian Science Fund (FWF) 'NEqD-si1D' [DOI: 10.55776/P35390], the DFG/FWF CRC 1225 ’ISOQUANT’ [DOI: 10.55776/I4863], the ERC-AdG Emergence in Quantum Physics (EmQ) [Grant Agreement No. 101097858], the ESPRIT grant ``Entangled Atom Pair Quantum Processor'' [DOI: 10.55776/ESP310], grant NSF PHY-1748958 to the Kavli Institute for Theoretical Physics (KITP), and NSF PHY-2210452
to the Aspen Center for Physics

\newpage

\section*{Supplementary Material} \label{section:Supplementary}
\renewcommand\thefigure{S\arabic{figure}}
\renewcommand\theequation{S\arabic{equation}}
\setcounter{equation}{0}

\subsection{Experimental system}
The experiments are performed with Feshbach molecules of $^6$Li. Our experimental setup and the preparation procedure have been described in detail in \cite{Liang_2022_mBEC_diffraction}.

In this work, the experimental cycle prepares a two-state mixture of lithium atoms in the hyperfine states $\ket{ F = 1/2,m_F = 1/2 }$ (state $\ket{1}$) and $\ket{ F = 3/2,m_F = -3/2 }$ (state $\ket{3}$). Then by evaporative cooling in a single beam dipole trap on the BEC side (675 G) of the 690 G Feshbach resonance \cite{PhysRevLett.94.103201} the atoms form weakly bound Feshbach molecules each consisting of two atoms in different hyperfine states. With further evaporation, the molecular cloud subsequently form a molecular BEC (mBEC) \cite{Jochim_2003_mBEC}.

The procedure prepares near pure mBECs containing up to $\sim$ 10000 $^6\mathrm{Li}_2$ Feshbach molecules \cite{Jochim_2003_mBEC}. The $s$-wave scattering length between molecules is tuned by setting the magnetic field \cite{PhysRevLett.110.135301}. For weakly bound molecules close to the Feshbach resonance, the dimer-dimer $s$-wave scattering length is given by $a_{dd} = 0.6 a_{13}$ \cite{PhysRevLett.93.090404}, where $a_{13}$ is the scattering length between atoms in states $\ket{1}$ and $\ket{3}$.

A qualitative sketch of the experimental arrangement can be seen in Fig.~\ref{fig:setup_and_profiles}(a). The mBEC is confined in a trapping potential formed by a focused laser beam (the capture beam) and the magnetic field curvature produced by the electric coils. The combined potential provides trap frequencies $(f_x,f_y,f_z) = {(16,100,100)}\,\mathrm{Hz}$, where $x$ denotes the axial direction along the trapping beam, $y$ the other horizontal direction, and $z$ the vertical (gravitational) direction. The field curvature is confining horizontally and therefore enhances trapping along the axial direction. Over the range of magnetic fields used, the trap frequencies are only weakly affected by the change of field level. The axial trap frequency is varied by 3\%, for magnetic field offset from $615\,\mathrm{G}$ to $660\,\mathrm{G}$, while radial confinement is dominated by the optical dipole trap.

\subsection{Optical speckle potential}
The optical speckle is formed with 646 nm laser light produced through sum-frequency-generation (SFG) from 1110 nm and 1550 nm laser sources in a periodically poled lithium niobate (PPLN) crystal, using a holographic diffuser (Edmund Optics, DIFFUSER HOLO 5 DEG 25MM). The speckle light is blue detuned and therefore repulsive to the lithium molecules.

The speckle pattern is projected along either one of the radial directions ($y$ or $z$), perpendicular to the long axis ($x$) of the condensate. The scaling results shown in Figs~\ref{fig:Evolution} and \ref{fig:Exponents} were obtained by pointing the speckle pattern along the (horizontal) $y$ direction, resulting in excitations along $x$ and $z$ directions, as shown in Fig.~\ref{fig:setup_and_profiles}(c). With this setup, the radial direction perpendicular to the speckle plane is imaged. In order to check for possible dynamics along the other radial direction which is in the plane of the speckle pattern, we took measurements also with the speckle pattern projected along the $z$ direction. No notable dynamics is observed in the transverse direction (Fig.~\ref{Fig:SpeckleSetup}(c)), while consistent results for scaling behaviours are obtained with the two setups. Additionally, the measurements with two sequential speckle pulses were obtained with the speckle projected vertically, as shown in Fig.~\ref{fig:scaling_vertical}.

Fig.~\ref{Fig:SpeckleSetup}(a) depicts the two alternative setups. In both cases, the diffused light after the diffuser is collected and shaped by focusing lenses, and consequently projected onto the condensate. For the vertical speckle setup, the imaging beam and speckle light are counter-propagating, with the paths combined at a dichroic mirror (CHROMA, F48-647SG) below the imaging objective (Special Optics 57-28.1-28) (Fig.~\ref{Fig:SpeckleSetup}(b)), which is the final optics projecting the speckle.  The characteristic grain size could be changed by adjusting the iris size. After optimizing the preparation procedure (see Supplementary: Quench sequence and breathing suppression), we chose an iris opening diameter of $15\,\mathrm{mm}$, estimated to give a characteristic grain size of $1\,\mathrm{\mu m}$ for the speckle potential, such that the cloud is excited enough to completely destroy the initial condensate, while not overheated so that no condensate could be formed again. For the horizontal ($y$) speckle setup, the final lens is a $65\,\mathrm{mm}$ achromat.


\subsection{Quench sequence and breathing suppression}

Experimental studies with a strongly interacting system are confronted with a major challenge, that a quench operation normally leads to breathing oscillations of the cloud, which complicates the system’s dynamics and also hinders the analysis of the momentum profile. The idea of our experimental approach is to introduce a broad range of excitations to a condensate by pulsing an optical speckle potential. However directly applying a speckle pulse similarly leads to breathing. To prevent this, we prepare the non-equilibrium state by injecting energy into the system with an empirically optimized procedure, which effectively diminishes breathing. First, an additional laser beam perpendicular to the trapping beam with similar radial trapping frequency ($f_r\sim 100\mathrm{Hz}$) is ramped up in 2ms and then abruptly switched off, causing the mBEC to axially ($x$) contract and subsequently expand to approximately double the original size. At the time point when the cloud expands to the maximum axial size, the speckle potential pulse is applied, rendering a broad range of excitations. The optimal time point is found via monitoring the breathing motion through Fourier transform of the evolution. This optimized procedure results in a monotonic evolution of the longitudinal momentum profile following the speckle pulse, for which we can apply the scaling analysis. Fig.~\ref{Img:BreathingCancel} shows the in-situ measurement of the axial cloud size with and without the contract-expand process. It can be clearly seen that the breathing motion is suppressed with the process. This approach is found to work for molecule numbers $N \geq 4000$.

Similarly, the squeeze-expand procedure could be replaced by an additional speckle pulse. After a direct speckle pulse on BEC, a second speckle pulse is applied when the cloud expand to the maximum size of the breathing mode. This double speckle scheme could also result in a monotonic evolution of momentum, as shown in Fig.~\ref{fig:scaling_vertical}. With double speckle scheme, the 1D scaling window is found to be absent, while the same exponent ratio $\alpha/\beta\approx0.5$ and duration of universal dynamics are observed for the crossover window.


\subsection{Momentum measurement}
Momentum measurement is realized with matter wave focusing \cite{Murthy_2014_focusing}. In our experiment, the capture beam is switched off abruptly, while the magnetic coils are kept on. The magnetic field curvature provides a focusing potential in the horizontal directions during the time-of-flight (TOF). The cloud expands rapidly along the radial directions, hence quickly reducing interaction. After a quarter of the oscillation period set by the horizontal trapping frequency of $16\,\mathrm{Hz}$, the spatial profile along the axial ($x$) direction corresponds to the momentum distribution prior to trap release. Detection is done with absorption imaging, integrating over the vertical ($z$) direction.

The momentum distributions in the analysis are obtained by folding the observed profile around the center point. The $k=0$ center point is determined by fitting the profile to a concentric-double-Gaussian function.


\subsection{Scaling Analysis}
\label{app:likelihood}

Without assuming the form of the scaling function $F_S[ \left(t/t_0\right)^{\beta}\,z ]$ (equation \ref{equation:scaling_evolution}), we find the time windows of scaling evolution and determine the exponents $\alpha$ and $\beta$ by computing the squared deviation of the scaled profiles from the profile of reference time $t_0$,

\begin{eqnarray}
    \chi^2(\alpha,\beta) &=& \frac{1}{N_t^2}\sum_{t,t_0}^{N_t}\chi_{\alpha,\beta}^2(t,t_0) \quad, \\[10pt]
    \chi_{\alpha,\beta}^2(t,t_0) \! &=& \!\!\! \int \! \frac{\left[ \tau^{\alpha}\,n(\tau^{\beta}k,t_0) - n(k,t) \right]^2}{\tilde{\sigma}(\tau^{\beta}k,t_0)^2 + \tilde{\sigma}(k,t)^2}\,dk ~, 
\end{eqnarray}
where $\tau=(t/t_0)$, and $\tilde{\sigma}(k,t)$ denotes the standard error of the mean of the profiles. The squared deviation for any particular pair of values of the exponents $\alpha$, $\beta$ measures how much the scaled profiles deviate from each other, therefore the summed deviation indicates how well the scaled profiles mutually agree. The calculation of deviation is performed from $k=0$ up to a scaling momentum cut-off $k_{cutoff}$.

In order to identify the time windows where the system exhibits universal scaling, we set a relatively small momentum cut-off and a narrow analysis time window, then scan the window through time, calculating the ratio of deviation between the rescaled profiles and that of the raw profiles $\chi^2_{0,0}/\chi^2_{\alpha,\beta}$. Fig.~\ref{fig:scalingWindow}(a) shows an example of the analysis with $k_{cutoff}\sim0.5/\mathrm{\mu m}$ and time window $t=5\mathrm{ms}$, where each peak reveals the possible existence of scaling dynamics. Then by scanning the momentum cut-off and analysis time window, we look for the maximum range that returns a value of $\chi^2_{0,0}/\chi^2_{\alpha,\beta}$ above 0.7 of the maximum value, and a scaling window is confirmed.

After determining the scaling window and momentum cut-off, the scaling exponents and the estimation of their uncertainties are then obtained from the position and the widths of the peak of the likelihood function.

\begin{equation}
    L(\Delta_{\alpha \beta},\beta) = exp\left[ -\frac{1}{2}\chi^2(\Delta_{\alpha \beta},\beta) \right] \quad.
\end{equation}

To examine the robustness of the result, the procedure for the scaling exponents' determination is also carried out by taking each holding time as the reference time separately. Fig.~\ref{fig:likelihood} shows the exponents found, which agree within the error bars given by the widths of the likelihood functions.

To see more clearly how the system out of equilibrium enters and leaves the scaling regime, we calculate the global observables $\bar{N}$ and $\bar{M}_2$ at each holding time \cite{Erne_2018_universal_scaling}.

\begin{eqnarray}
    \bar{N} &=& \int_{\left| k \right| \leq (t/t_0)^{-\beta}k_{cutoff}}\,\frac{n(k,t)}{N(t)}\,dk \quad, \\[10pt]
    \bar{M}_{n \geq 2} &=& \int_{\left| k \right| \leq (t/t_0)^{-\beta}k_{cutoff}}\,\frac{\left| k \right|^2 n(k,t)}{N\bar{N}(t)}\,dk \quad,
\end{eqnarray}
The observable $\bar{N}$ gives the fraction of particles in the scaling region, whereas $\bar{M}_2$ is related to the mean kinetic energy per particle in the scaling region. The results of the calculation if shown in Fig.~\ref{fig:globalvariable}.

\subsection{Random Defect Model}
The random defect model (RDM) \cite{Schmidt_2012_RDM} describes the momentum distribution of a dilute ensemble of solitonic defects. In a uniform background, it has the form:

\begin{equation}
    n(k)=\left[\frac{2n_{0}k_{n_s}}{k_{n_s}^2+(k-k_{0})^2}\right]\left[ \frac{k/k_{\xi_s}}{\mathrm{sinh}(k/k_{\xi_s})} \right]^2 \quad ,
\end{equation}

where $n_0$ is the uniform background density, $k_0$ is the bulk momentum of the soliton ensemble, $k_{n_s}=2n_s$ is the mean distance between solitons with $n_s$ being the soliton density and $k_{\xi_s}=\sqrt{2}/(\pi \xi_s)$ is the momentum scale related to the width $\xi_s$ of single soliton.

We first determine $k_{\xi_s}$ by fitting the exponential range of the distribution (high momentum range), where the distribution could be approximated by $n(k)\propto \mathrm{exp}(-2\frac{k}{k_{\xi_s}})$. Afterward, we fix the value of $k_{\xi_s}$ and fit the full function to obtain the soliton density $n_s$ (Fig~\ref{fig:solitonDensityFit}). $k_0$ in the distribution is taken to be zero.

\subsection{Universal Function}
Shown in Fig.~\ref{fig:Exponents}(c)(d), all folded profiles within the ``1D'' (crossover) scaling window collapse to a single universal function $f_s=1/(1+\abs{k}^\zeta)^{-1}$ with exponent $\zeta=1.604\pm0.011(1.766\pm0.014)$. The value of $\zeta$ deviate from the expected value of fitting to the theory distribution of RDM. When an offset term is introduced to the universal function $f_s\sim\left[1+(\abs{k}+k_{int})^{\zeta} \right]^{-1}$, the value of $\zeta$ change from 1.6 to 2.56 with a fitted offset of $k_{int}\approx0.05\mu m^{-1}$, matching the prediction (Fig~\ref{fig:RDMfitting}(a)). 

The additional factor of $k_{int}$ could be explained by the release of interaction energy during matter wave focusing, which constitute $\sim 1.5\%$ of the total chemical potential. When the cloud is released from the dipole trap, most of the interaction energy is released into the transverse dimensions ($y$,$z$). However, a small fraction of the interaction energy would still be released in the longitudinal direction ($x$), causing an overall increase in the observed momentum distribution. As shown in Fig~\ref{fig:RDMfitting}(b), by calculating the difference between the theoretical model and the measured distribution, an increasing trend with interaction is observed.

\subsection{Crossover Scaling Window and Transverse Trapping}
In the second scaling window, we observe scaling exponents of $\alpha\approx\frac{1}{2}\beta$, with the time window relatively stable versus interaction strength. Interestingly, both the starting time and the duration of the second window are on the order of inverse transverse trap frequency ($2\pi/\omega_{\perp}$). We perform additional measurements with different trapping geometry ($f_{\parallel}=16\,\mathrm{Hz}$, $f_{\perp}=56\,\mathrm{and}\,200\,\mathrm{Hz}$), and the results (Fig~\ref{fig:scalingCossoverWindow}) confirm the connection of second scaling window to transverse dynamic. So we named it the ``crossover'' scaling window.

\newpage
\appendix


\onecolumngrid
\renewcommand\thefigure{S\arabic{figure}}
\setcounter{figure}{0}
\renewcommand\thetable{S\arabic{table}}
\setcounter{table}{0}

\begin{figure}[!h]
\center
\includegraphics[width=0.6\columnwidth]{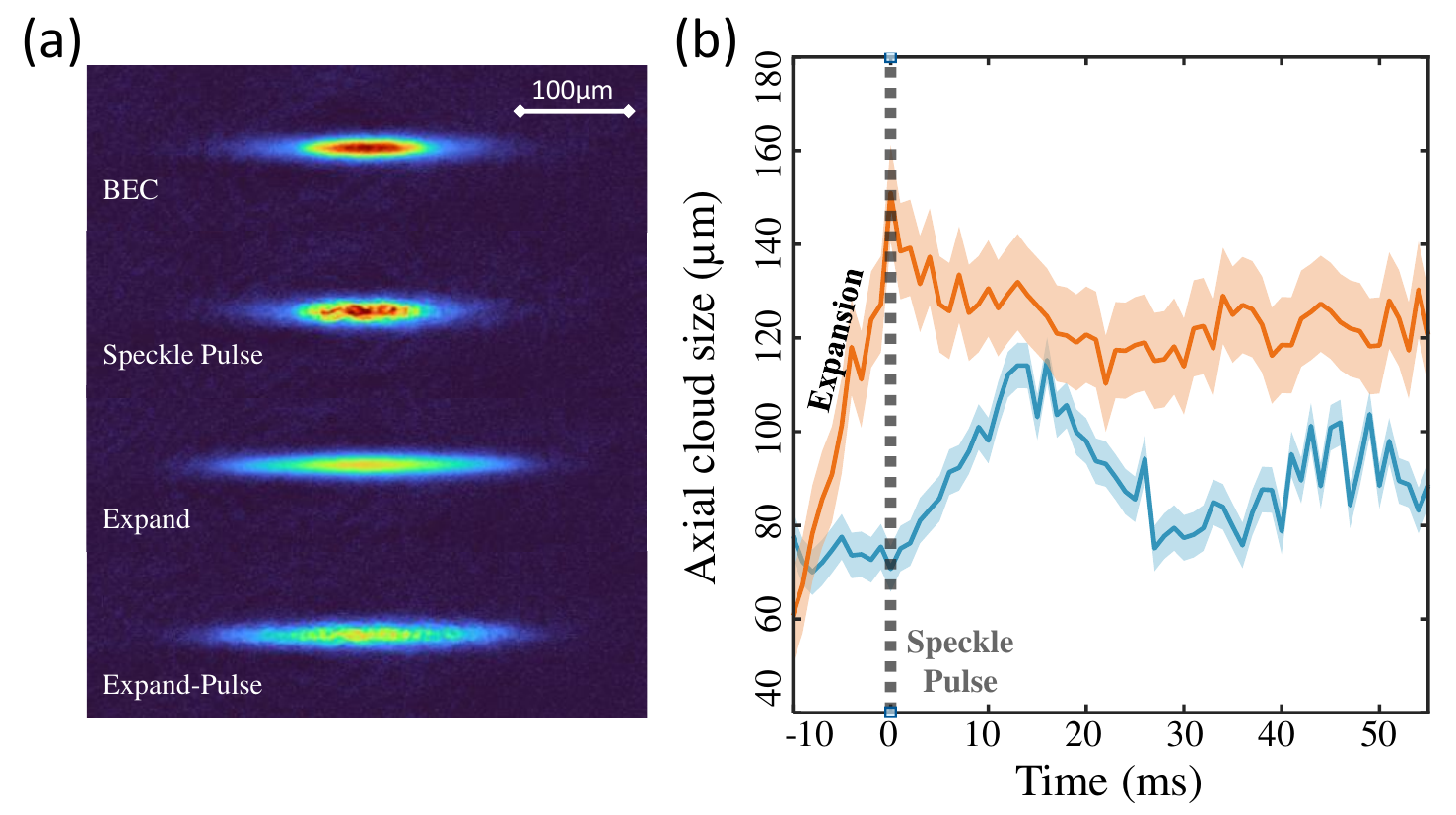}
\caption{\textbf{In-situ density profile and size measurement:} (\textbf{a}) In-situ density profile of the initial condensate and immediately after different procedures. (\textbf{b}) In-situ cloud size measurement after speckle pulse with and without the breathing suppress procedure.}
\label{Img:BreathingCancel}
\end{figure}

\begin{figure}[!h]
    \centering
    \includegraphics[width=0.8\columnwidth]{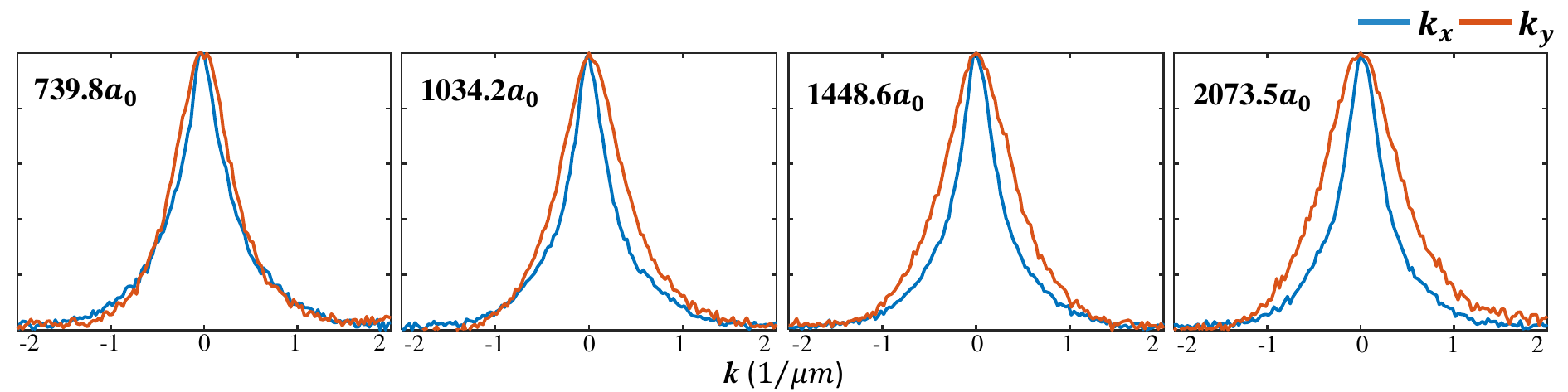}
    \caption{\textbf{Momentum distribution of initial state at different interaction:} Momentum distribution of the initial state ($t=0$) in the axial and radial directions, measured for different interactions. The width of the radial momentum distribution increases with stronger interaction while the axial distribution remains essentially unchanged, indicating that most of the interaction energy is released into the radial directions during TOF expansion. This provides the check of validity for our momentum measurements in the axial direction.}
    \label{fig:transversMomentum}
\end{figure}

\begin{figure}[!h]
\center
\includegraphics[width=0.9\columnwidth]{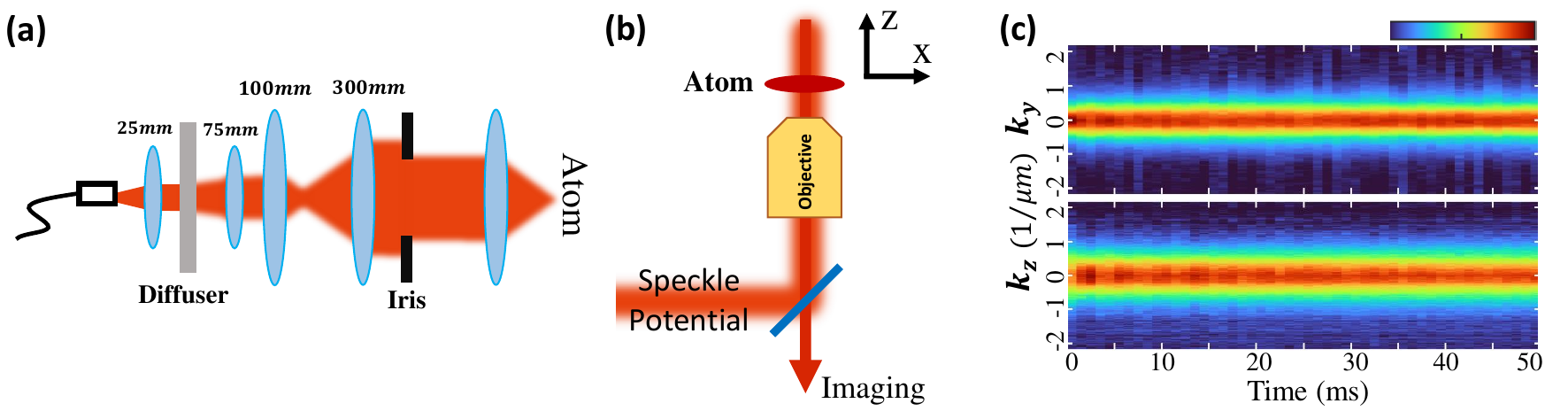}
\caption{\textbf{Speckle beam setup and transverse momentum measurement:} Schematics of the horizontal (\textbf{a}) and the vertical (\textbf{b}) speckle setups, with last lens being a $65\mathrm{mm}$ achromat, or the objective of the imaging system, respectively. (\textbf{c}) Transverse momentum measured with both horizontal and vertical speckle setup.}
\label{Fig:SpeckleSetup}
\end{figure}

\begin{figure}[!h]
\center
\includegraphics[width=0.5\columnwidth]{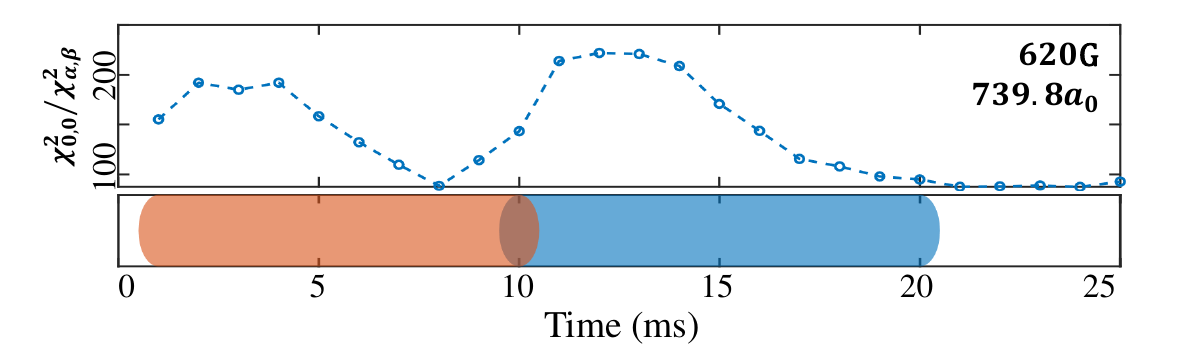}
\caption{\textbf{Scaling windows determination:} $\chi^2_{0,0}/\chi^2_{\alpha,\beta}$ calculation and Scaling window analysis results for $B=620\,\mathrm{G}$, $a_{dd}=739.8a_0$. Two distinct peaks suggest two separate scaling regimes.}
\label{fig:scalingWindow}
\end{figure}

\begin{figure}[!h]
\center
\includegraphics[width=0.5\columnwidth]{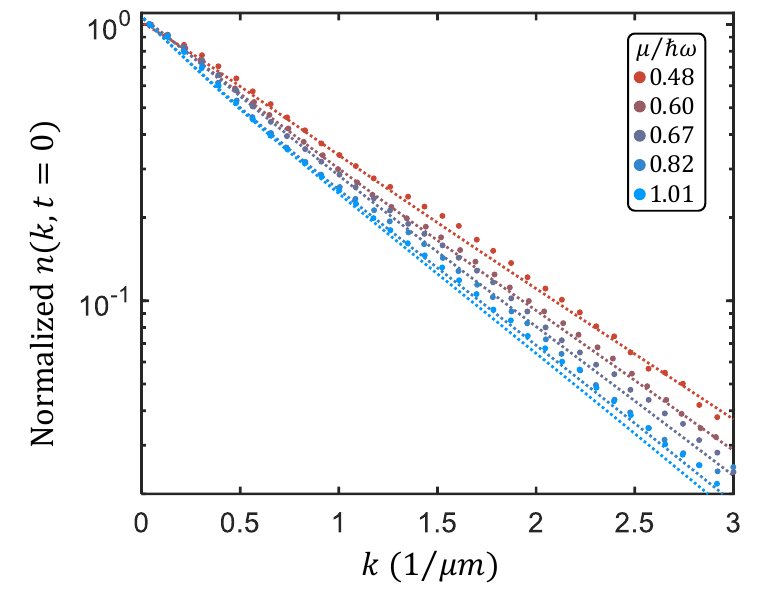}
\caption{\textbf{Fitting initial state momentum:} Initial momentum distribution $n(k,t=0)$ directly following the quench. Solid lines are the fit result of the random defect model. Data points are plotted with binning of 3.}
\label{fig:solitonDensityFit}
\end{figure}

\begin{figure}[!h]
\center
\includegraphics[width=0.8\columnwidth]{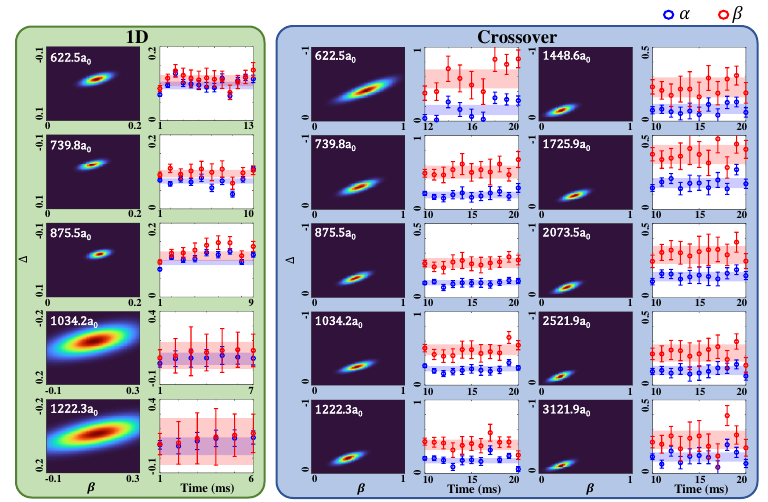}
\caption{\textbf{Likelihood calculation and robustness check:} The calculation of likelihood function and exponent robustness check of both scaling windows.}
\label{fig:likelihood}
\end{figure}

\begin{figure}[!h]
\center
\includegraphics[width=0.8\columnwidth]{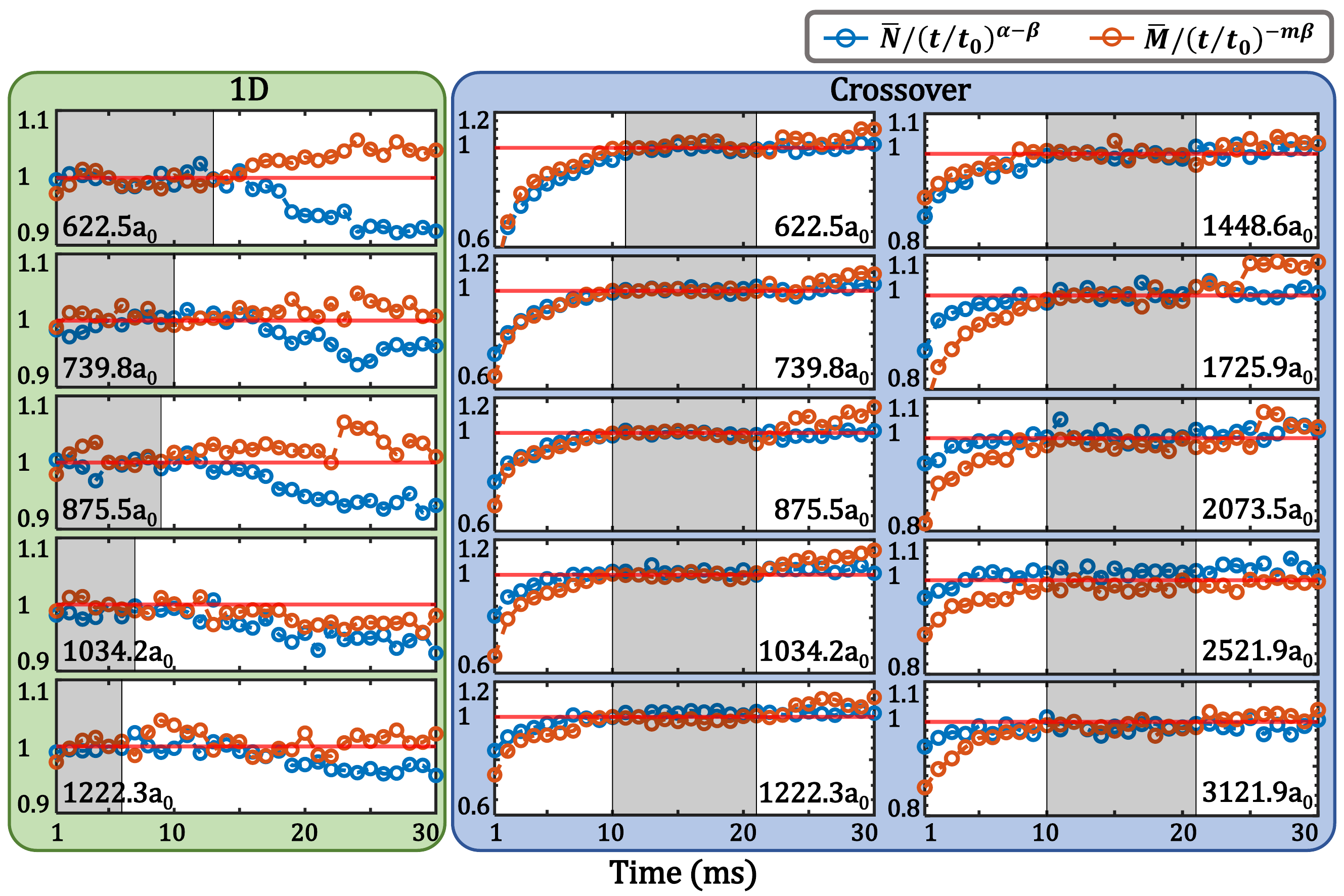}
\caption{\textbf{Global variable calculations:} Global variable calculations for data sets measured with particle number 7000$\sim$10000 and scattering length 622$\sim$3122$a_0$. The values of $\overline{N}$ and $\overline{M}$ are rescaled by the respective exponential functions, such that their values equal to 1 (marked by the horizontal solid line) when the system is in the scaling regime (marked by the gray area).}
\label{fig:globalvariable}
\end{figure}

\begin{figure}[!h]
\center
\includegraphics[width=0.8\columnwidth]{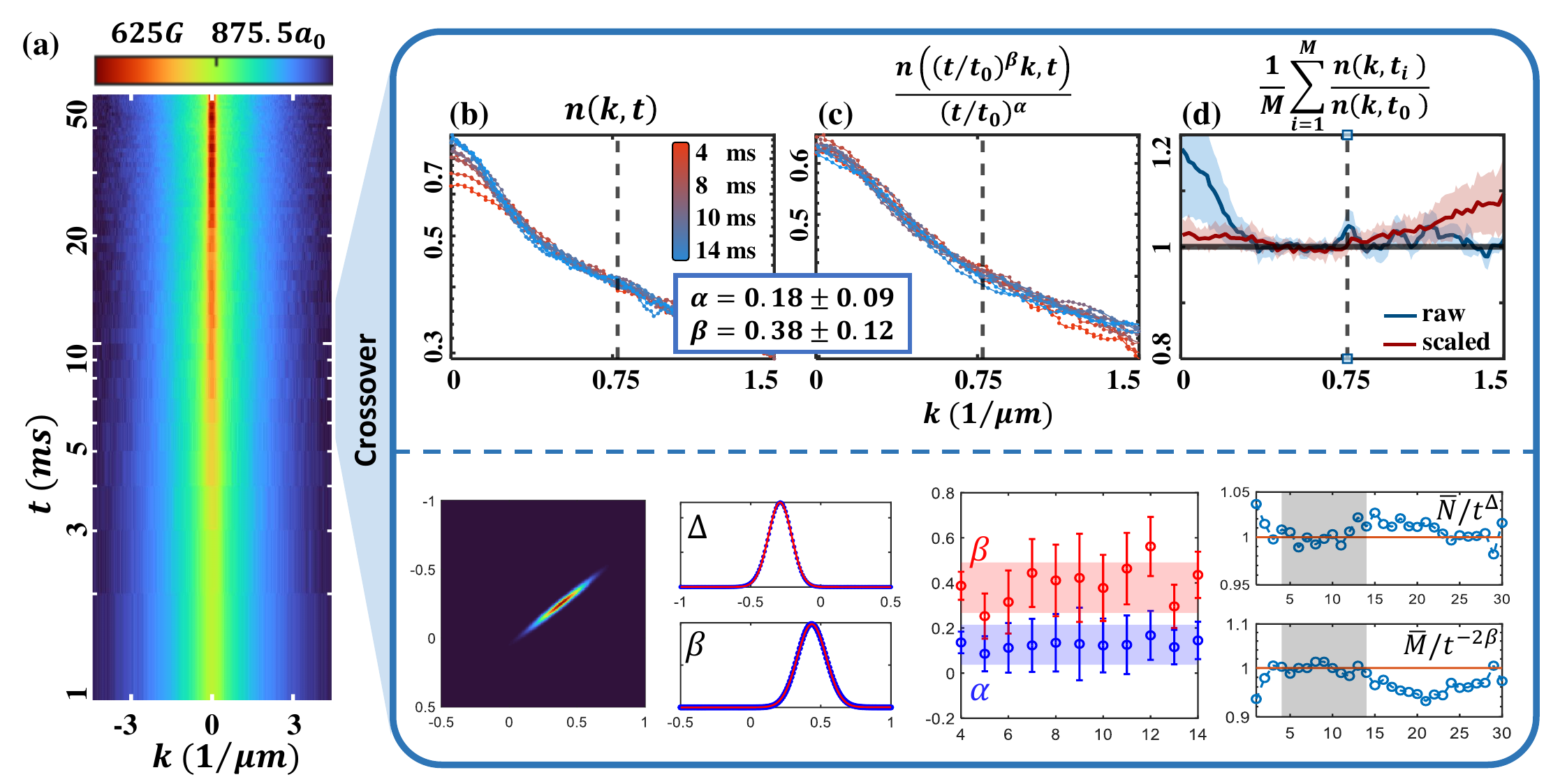}
\caption{\textbf{Double speckle sequence measurement:} Momentum evolution measured with $B=625\,\mathrm{G} , a=876a_{0}$ using double speckle sequence. (\textbf{a}) The time carpet of the transverse momentum distribution. The (\textbf{b})raw and (\textbf{c})scaled momentum profiles. (\textbf{d})The mean ratio of raw (blue) and scaled (red) profile divided by the reference profile. Scaling cutoff is marked by vertical dashed lines.}
\label{fig:scaling_vertical}
\end{figure}

\begin{figure}[!h]
    \centering
    \includegraphics[width=0.8\columnwidth]{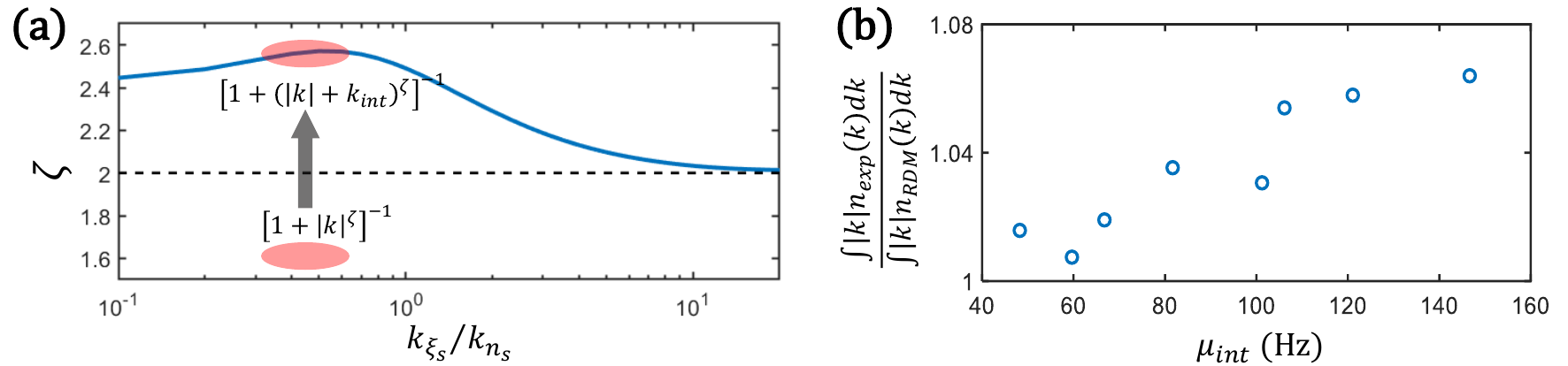}
    \caption{\textbf{Fitting universal function:} (\textbf{a}) Value of the exponent $\zeta$ (solid line) by fitting to the theoretical distribution $[1+(\frac{k_{\xi_s}}{k_{n_s}})^2(\frac{k}{k_{\xi_{s}}})^2]^{-1}$, and comparison to the fitting value of the experimental data (red area). (\textbf{b}) The ratio of the total momentum $\int|k|n(k)dk$ between the theoretical model and measured distribution.}
    \label{fig:RDMfitting}
\end{figure}

\begin{figure}[!h]
    \centering
    \includegraphics[width=0.5\columnwidth]{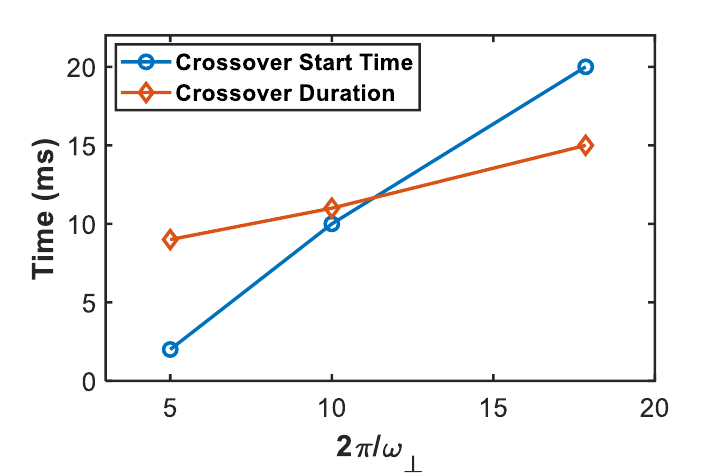}
    \caption{\textbf{Crossover window:} Starting time and duration of "crossover" scaling window plot versus the inverse trap frequency.}
    \label{fig:scalingCossoverWindow}
\end{figure}

\begin{table}[h]
\center
\begin{tabular}{|p{25pt}|p{25pt}|p{30pt}|p{20pt}|p{20pt}|p{40pt}|p{43pt}|p{20pt}|p{40pt}|p{43pt}|p{20pt}|p{30pt}|p{40pt}|}
 \hline
 \multicolumn{13}{|c|}{\textbf{Lithium data} ($^6$Li$_2$), $(\nu_\parallel,\nu_\perp) = (16,100) \, \mathrm{Hz}$} \\
 \hline
 $a_s$ & $N$ &$\mu_{int}$ & $\xi_h$ & $T_\mathrm{final}$& \multicolumn{3}{|c|}{1D-scaling} & \multicolumn{3}{|c|}{crossover-scaling}  & $\xi_s$& $n_s$ \\
  $[a_0]$& &$[h \, \mathrm{Hz}]$& $[\mu \mathrm{m}]$& $[\mathrm{nK}]$ & $\alpha$ & $\beta$ & $t[\mathrm{ms}]$ & $\alpha$ & $\beta$ & $t[\mathrm{ms}]$ &$[\mu \mathrm{m}]$ & $[\mu \mathrm{m}^{-1}]$ \\
 \hline
   623& 6777&   49.99& 2.90& 43.2&    0.095(11)& 0.108(15)& 13&      0.152(96) & 0.574(128) & 9 &       1.29(2) & 0.483(15)  \\
   740& 7500&   60.51& 2.64& 43.8&    0.075(7)& 0.095(10)& 10&      0.194(45) & 0.504(87) & 10 &       1.41(3) & 0.403(22)  \\
   876& 7853&   67.67& 2.50& 43.6 &    0.093(7) & 0.111(13) & 9 &      0.198(36) & 0.466(69) & 10 &       1.48(3) & 0.394(14)  \\
   1034& 9267&   83.05& 2.25& 46.1&   0.076(42) & 0.099(92) & 7 &      0.223(35) & 0.483(74) & 10 &       1.61(3) & 0.291(68)  \\
   1222& 9184&   103.41& 2.02& 45.7&    0.081(61) & 0.114(159) & 6 &      0.171(37) & 0.389(71) & 10 &       1.65(3) & 0.287(53)  \\
   1449& 9300&   108.67& 1.97& 46.4&    -- & -- & -- &      0.083(34) & 0.228(68) & 10 &       1.65(3) & 0.372(22)  \\
   1726& 9345&   123.90& 1.84& 46.9&    -- & -- & -- &      0.178(34) & 0.371(66) & 10 &       1.70(3) & 0.302(48)  \\
   2074& 9323&   150.14& 1.68& 46.9&    -- & -- & -- &      0.142(32) & 0.289(62) & 10 &       1.72(3) & 0.361(12)  \\
   \hline
   2522& 9553&   --& --& 47.1&    -- & -- & -- &      0.099(32) & 0.219(59) & 10 &       1.72(3) & 0.360(07)  \\
   3122& 10063&   --& --& 48.4&    -- & -- & -- &      0.111(32) & 0.228(63) & 10 &       1.85(4) & 0.326(08)  \\
 \hline
\end{tabular}

\vspace{0.5cm}
\begin{tabular}{|p{25pt}|p{25pt}|p{30pt}|p{30pt}|p{30pt}|p{30pt}|p{30pt}|p{30pt}|p{30pt}|p{30pt}|p{30pt}|p{35pt}|p{40pt}|}
 \hline
 \multicolumn{13}{|c|}{\textbf{Rubidium data} ($^{87}$Rb), $(\nu_\parallel,\nu_\perp) = (23,3300) \, \mathrm{Hz}$} \\
 \hline
 $a_s$ & $N$ &$\mu_{int}$ & $\xi_h$ & $T_\mathrm{final}$& \multicolumn{3}{|c|}{1D-scaling} & \multicolumn{3}{|c|}{crossover-scaling}  & $\xi_s$& $n_s$ \\
  $[a_0]$& &$[h \, \mathrm{Hz}]$& $[\mu \mathrm{m}]$& $[\mathrm{nK}]$ & $\alpha$ & $\beta$ & $t[\mathrm{ms}]$ & $\alpha$ & $\beta$ & $t[\mathrm{ms}]$ &$[\mu \mathrm{m}]$ & $[\mu \mathrm{m}^{-1}]$ \\
 \hline
  98 & 1700 &    1099 & 0.23 &   100 &        0.09(4) & 0.10(4) & 75 &      -- & -- & -- &   0.07 & 1.4 \\
  98 & 2800 &    1454 & 0.20 &   -- &        0.09(6) & 0.10(7) & 37 &      -- & -- & -- &   0.06 & 0.9 \\
  98 & 1150 &    691 & 0.29 &   -- &        0.09(4) & 0.11(5) & 63 &      -- & -- & -- &   0.05 & 2.3 \\
 \hline
\end{tabular}
\caption{
\textbf{Datasets overview for Lithium and Rubidium data:} The momentum measurements were made with interaction strength $620a_0$ to $3100a_0$. In situ density was only measured for interaction strength $620a_0$ to $2000a_0$. The Rubidium data is taken from \cite{Erne_2018_universal_scaling}.}
\label{Table:Dataset_overview}
\end{table}

\clearpage
\twocolumngrid
\bibliography{Li6_universal}


\end{document}